\documentclass{elsart3p}
\journal{Phys. Lett. B}
\usepackage{latexsym}
\usepackage{epsfig}
\usepackage{amssymb}
\newcommand{\beq}{\begin{eqnarray}}
\newcommand{\eeq}{\end{eqnarray}}
\newcommand{\be}{\begin{eqnarray}}
\newcommand{\ee}{\end{eqnarray}}
\newcommand{\bk}{{\bf k}}

\newcommand{\bx}{{\bf x}}

\begin{document}
\begin{frontmatter}
\title{Cosmology and Astrophysical Constraints of Gauss-Bonnet Dark Energy}
\author[TK]{Tomi Koivisto\ead{tomikoiv@pcu.helsinki.fi}}
\and
\author[DFM,DFMM]{David F. Mota\ead{mota@astro.uio.no}}
\address[TK]{Helsinki Institute of Physics,FIN-00014 Helsinki, Finland}
\address[DFM]{ Institute of Theoretical Astrophysics, University of Oslo, 0315 Oslo, Norway}
\address[DFMM]{Perimeter Institute for Theoretical Physics, Waterloo, Ontario N2L 2Y5, Canada}
\begin{abstract}
Cosmological consequences of a string-motivated dark energy scenario 
featuring a scalar field coupled to the Gauss-Bonnet invariant are investigated.
We study the evolution of the universe in such a model, identifying 
its key properties. The evolution 
of the homogeneous background and cosmological perturbations, both at large and small 
scales, are calculated. The impact of the coupling on galaxy 
distributions and the cosmic microwave background is examined.  
We find the coupling provides a mechanism to viably onset the late acceleration, to alleviate the coincidence problem,
and furthermore to effectively cross the phantom divide at the present while avoiding
a Big Rip in the future. We show the model could explain the present
cosmological observations, and discuss how various astrophysical and cosmological data,
from the Solar system, supernovae Ia, cosmic microwave background radiation and large 
scale structure constrain it. 
\end{abstract}
\end{frontmatter}
Einstein General Relativity together with ordinary matter,
described by the standard model of particle physics, 
cannot fully explain the observational data
from Supernovae type Ia (SNeIa)  \cite{Riess:2004nr}, the matter 
power spectrum
of large scale structure(LSS)   \cite{sloan} and the anisotropy spectrum
of the Cosmic Microwave Background Radiation(CMBR)   \cite{Spergel:2006hy}. One
needs to introduce two exotic components into the matter-energy
budget of the Universe.  Dark matter, a fluid with zero or very small 
pressure, corresponds to about $25\%$ of the universe energy
budget.  Dark energy, with its negative pressure, dominates the Universe density 
and is responsible for its present acceleration  \cite{Copeland:2006wr}.

The observed acceleration and the undisclosed nature of these two exotic components strongly 
motivates the extension of both the standard model of particle and
gravitational physics sectors \cite{Nojiri:2006ri,Neupane:2006ip,Nojiri:2005jg}. 
This is also what quantum gravity seems to require. 
Typically the low-energy limit of string theory features scalar fields and
their couplings to various curvature terms. 
Interestingly, there is a unique combination of the curvature squared terms, the Gauss-Bonnet (GB) invariant
\be
R^2_{GB} \equiv R^{\mu\nu\rho\sigma}R_{\mu\nu\rho\sigma} - 4R^{\mu\nu}R_{\mu\nu} + R^2,
\ee
that is both ghost-free in Minkowski backgrounds and leads to second order order field 
equations. All versions of string theory in 10 dimensions (except Type II) include this term 
as the leading order $\alpha^\prime$ correction  \cite{Callan,gross}. 
Specifically, these couplings can be shown to appear at the tree level or the 
one-loop level (depending on whether one considers a dilaton or an average volume modulus) of  
string effective action when going from the string frame to the Einstein 
frame \cite{Antoniadis2,Antoniadis}.
The effective action could then be written as
\be S = \label{action}
\int d^4 x \sqrt{-g}\left[ \frac{R}{2\kappa^2} - f(\phi)R^2_{GB} + L_\phi + L_m\right],
\ee
where $\kappa=(8\pi G)^{-1/2}$. The scalar field Lagrangian 
is $L_\phi = -\frac{\gamma}{2}(\nabla\phi)^2 - V(\phi)$, where $\gamma$ is a constant.  The function
$f(\phi)=\sigma-\hat{\delta} \xi(\phi)$: the coupling $\sigma$ may be related to string 
coupling $g_s$ via $\sigma\sim 1/g_s^2$. The numerical coefficient $\hat{\delta}$ typically
depends on the massless spectrum of every particular model \cite{Antoniadis}.

In this work we explore the cosmological consequences and viability of the action (\ref{action}),
with specific choices for the potential and the coupling, both taken to be single exponential terms.  
Such models have been recently studied in \cite{Nojiri:2005vv} within the context of the dark energy problem,
and their possible background evolution has been investigated \cite{Carter:2005sd,Carter:2005fu,
Neupane:2006dp,Sami:2005zc,Calcagni:2005im,Nojiri:2006je}. 
We will show that even the simple and well-motivated 
exponential parameterization can naturally exhibit a viable transition to acceleration 
and a transient phantom expansion. 
Besides presenting such possibility in the model and quantitatively
testing it with various observational data, we also consider the phenomenology of GB dark energy
models in general at cosmological, astrophysical and Solar system scales during the whole cosmological 
evolution. In particular, we investigate how the coupling affects the CMBR anisotropies and LSS.

We consider a flat, homogeneous and 
isotropic background universe with scale factor $a(t)$. Derivatives with 
respect to the cosmic time $t$ are denoted by an overdot, and a prime 
 means derivative with respect to the e-folding time 
$\log{(a)}$ unless 
other variable is explicitly specified. Action (\ref{action}) yields 
the
Friedmann equation
\be \label{friedmann}
\frac{3}{\kappa^2}H^2 = \frac{\gamma}{2}\dot{\phi}^2+V(\phi) + \rho_m +
24H^3f'(\phi)\dot{\phi},
\ee
and the Klein-Gordon equation
\be \label{klein-gordon}
\gamma (\ddot{\phi}+3H\dot{\phi})+V'(\phi)+f'(\phi)R^2_{GB}=0,
\ee
where $H\equiv \dot{a}/a$ is the Hubble rate, $\rho_m$ represents the matter component, and the GB 
invariant is
$R^2_{GB}=24H^2(\dot{H}+H^2)$.
It is convenient to define the dimensionless variables
$\Omega_m \equiv \frac{\rho_m\kappa^2}{3H^2}$, 
$x \equiv \frac{\kappa}{\sqrt{2}}\frac{\dot{\phi}}{H}$, 
$y \equiv \kappa^2\frac{V(\phi)}{H^2}$, 
$\mu  \equiv 8\kappa^2\dot{\phi} H f'(\phi)$,
and
$\epsilon \equiv \frac{\dot{H}}{H^2}$.
We consider a canonical scalar field with $\gamma=1$ and
adopt an exponential form for the potential $V(\phi) = V_0 e^{-\lambda\kappa \phi/\sqrt{2}}$ and for the coupling $f = f_0 e^{\alpha\kappa \phi/\sqrt{2}}$.
The nonperturbative effects from gaugino condensation or instantons can 
result in an exponential potential \cite{Dine:1985rz}.
An exponential field-dependence can approximate the coupling ensuing from, for instance heterotic
compactification \cite{Antoniadis2,Antoniadis}. 

The background has altogether six fixed points when considered as a dynamical system.
However, only two of them are now relevant (see \cite{we} for details). 
The first is the  standard exponential tracking solution, where the scalar 
field mimics exactly the background equation of state $w_m$.
This fixed point is a stable spiral when $\lambda > \alpha, 
\sqrt{6}(1+w_m)$. It is, however, a saddle point when 
$\lambda,\alpha<\sqrt{6}(1+w_m)$ or 
$\alpha \ge \lambda,\sqrt{6}(1+w_m)$.
The later happens when the coupling becomes 
significant at late times. Then the field will be passed from the scaling solution to a 
potential-dominated solution  (see \cite{we}).
 
Since for the scaling solution we have $H^2 \sim \rho \sim a^{-3(1+w_m)}$, the last term in the Friedmann
equation scales like $\rho_f \equiv 8H^3f'(\phi)\dot{\phi} \sim a^{-3(1+w_m)(2-\alpha/\lambda)}$. This
follows from the tracking behaviour of the scalar field; since $\phi' =
3(1+w_m)/\lambda$, we have that $\phi = \phi_0 + 3(1+w_m)\log(a)/\lambda$,
and hence $f'(\phi) \sim a^{3\alpha(1+w_m)/\lambda}$. Thus we find that
the effective energy density due to the presence of the GB term, $\rho_f$,
dilutes slower than the energy density due to matter, $\rho_m$,
if and only if $\alpha>\lambda$. 

\begin{figure}
\begin{center}
\includegraphics[width=0.45\textwidth]{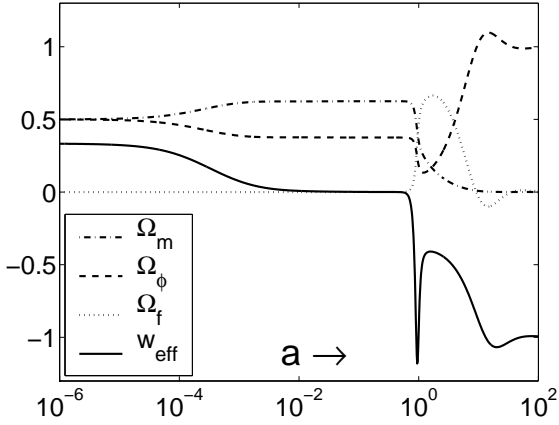}
\caption{\label{omegas} 
Background evolution when $\lambda=4$ and $\alpha=20$. 
In general, the evolution in the model consists of 1) the scaling 
attractor and 2) the potential-dominated (de Sitter) solution. Existence
of 1) requires $\lambda > \sqrt{6}$, and the transition to 2) then occurs 
if $\alpha>\lambda$. Here we plot fractional energy densities
for matter, $\Omega_m$ (dash-dotted line), scalar field
$\Omega_\phi = (x^2+y)/3$ (dashed line) and the GB
term, $\Omega_f = \mu$ (dotted line). Solid line is the total
equation of state $w_{eff} = -2\epsilon/3 - 1$.}
\end{center}
\end{figure}

A typical evolution for the model
is shown in FIG. \ref{omegas}. Note that $\Omega_\phi>1$ may occur, since the
(effective) GB energy density $\Omega_f$ can be negative 
, and that $w_{eff}<-1$ is possible here as well.

To investigate the cosmological effects of the GB coupling in more detail, 
we also consider the linear perturbations. In the synchronous gauge \cite{Ma:1995ey}, one
can parameterize the metric perturbations as
\be \nonumber
      d s^2 = -dt^2 + a^2(t)(\delta_{ij} + h_{ij}) dx^i dx^j.
\ee
Conventionally the scalar modes are then defined in the Fourier space by
the decomposition into the trace ($h$) and the traceless part ($\eta$) of $h_{ij}$,
\be \nonumber
       h_{ij}(\bx,\tau) = \int d^3k e^{i\bk \cdot \bx}
[\hat{k}_i\hat{k}_jh(\bk,\tau) + 6(\hat{k}_i\hat{k}_j -
       \frac{1}{3}\delta_{ij})\eta(\bk,\tau)],
\ee
where $\bk=k\hat{k}$. One can then write the energy constraint (perturbed version of the Friedmann equation) as an 
evolution equation for the metric potential $h$,
\be  
\dot{h} = \frac{8\pi G}{H(1-\frac{3}{2}\mu)}\left[2(1-\mu)\frac{\eta}{a^2}+\delta\rho + 8H^2(3H\dot{\delta f} 
+\frac{k^2}{a^2}\delta f)\right].
\ee
where the density fluctuation is as usual $\delta \rho = \rho_m\delta_m + \dot{\phi}\dot{\delta\phi}
+ V'(\phi)\delta\phi$. Note there appears both new source terms due to the fluctuations
in the coupling $f$ as well as a modulating prefactor changing the response to the standard
source terms\footnote{The expression implies divergence when $\mu = 2/3$ (unless it would happen 
that the square bracket term vanishes for all $k$-modes at that point). However, in the models considered
in this letter we always have the relative Gauss-Bonnet contribution $\mu<2/3$. Then also Eq.(\ref{e_eta}) is 
well-behaved.}.
Similar modifications are present in the momentum constraint equation governing
the evolution of the other metric potential $\eta$,
\be \label{e_eta}
\dot{\eta} = \frac{4\pi G}{1-\mu}\left[a(\rho+p)\theta - 8H(H\dot{\delta f}-H^2\delta f)\right],
\ee
where the velocity perturbation is as usual $a(\rho+p)\theta = a(\rho_m+p_m)\theta_m 
+ k\dot{\phi}\delta\phi$. The Klein-Gordon equation for the scalar field fluctuation, 
being the perturbed version of Eq.(\ref{klein-gordon}), is also non-trivially modified \cite{Hwang:2005hb}. Since 
still minimally coupled to gravity, all other matter obey the usual continuity 
equations \cite{we,Koivisto:2005yk}.

We have numerically integrated the fully perturbed 
equations \footnote{We used a modified version of the 
CAMB code \cite{Lewis:1999bs}.} 
and computed the full matter power and CMBR spectra for the example 
model presented here. 
\begin{figure}
\begin{center}
\includegraphics[width=0.35\textwidth]{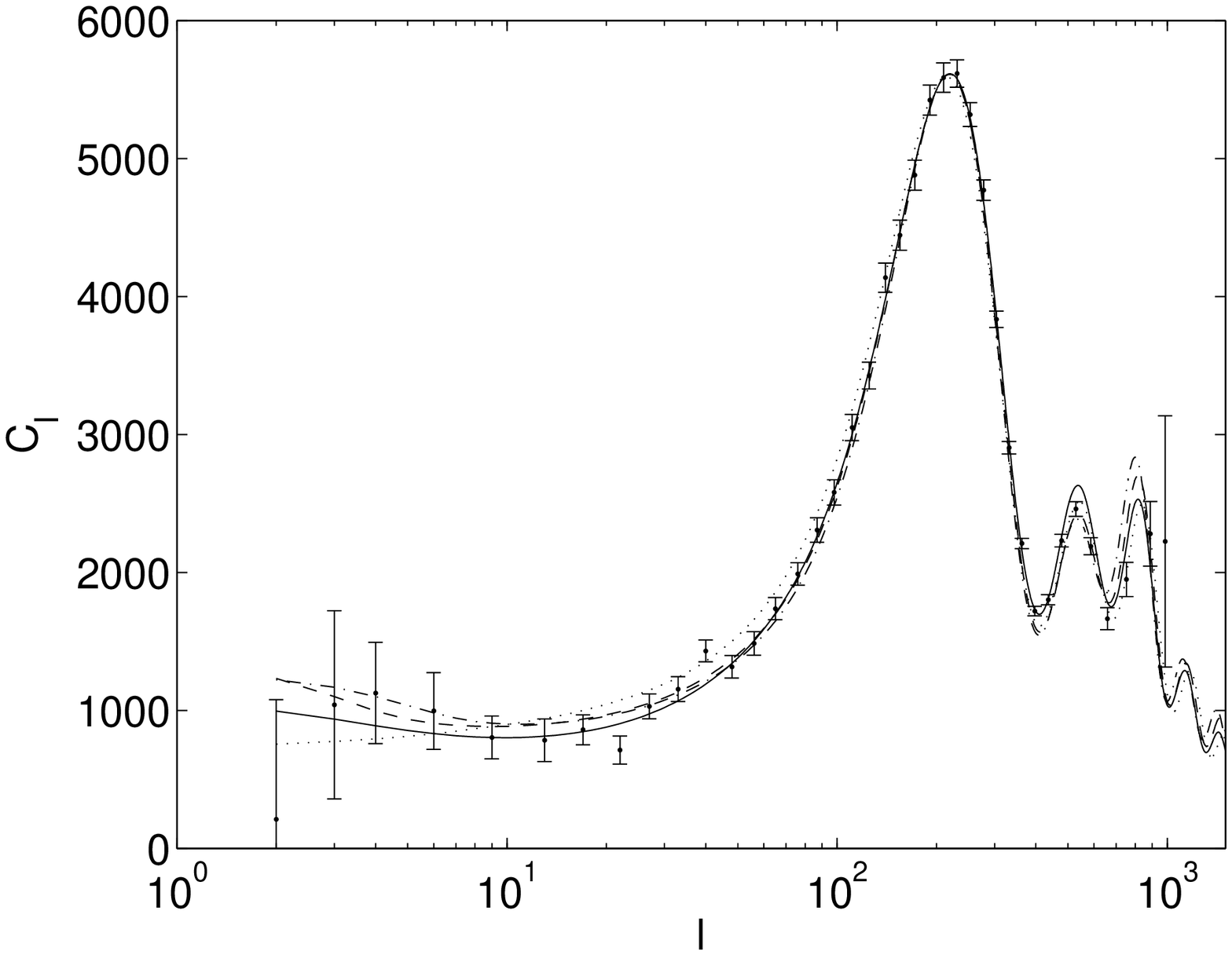}
\includegraphics[width=0.35\textwidth]{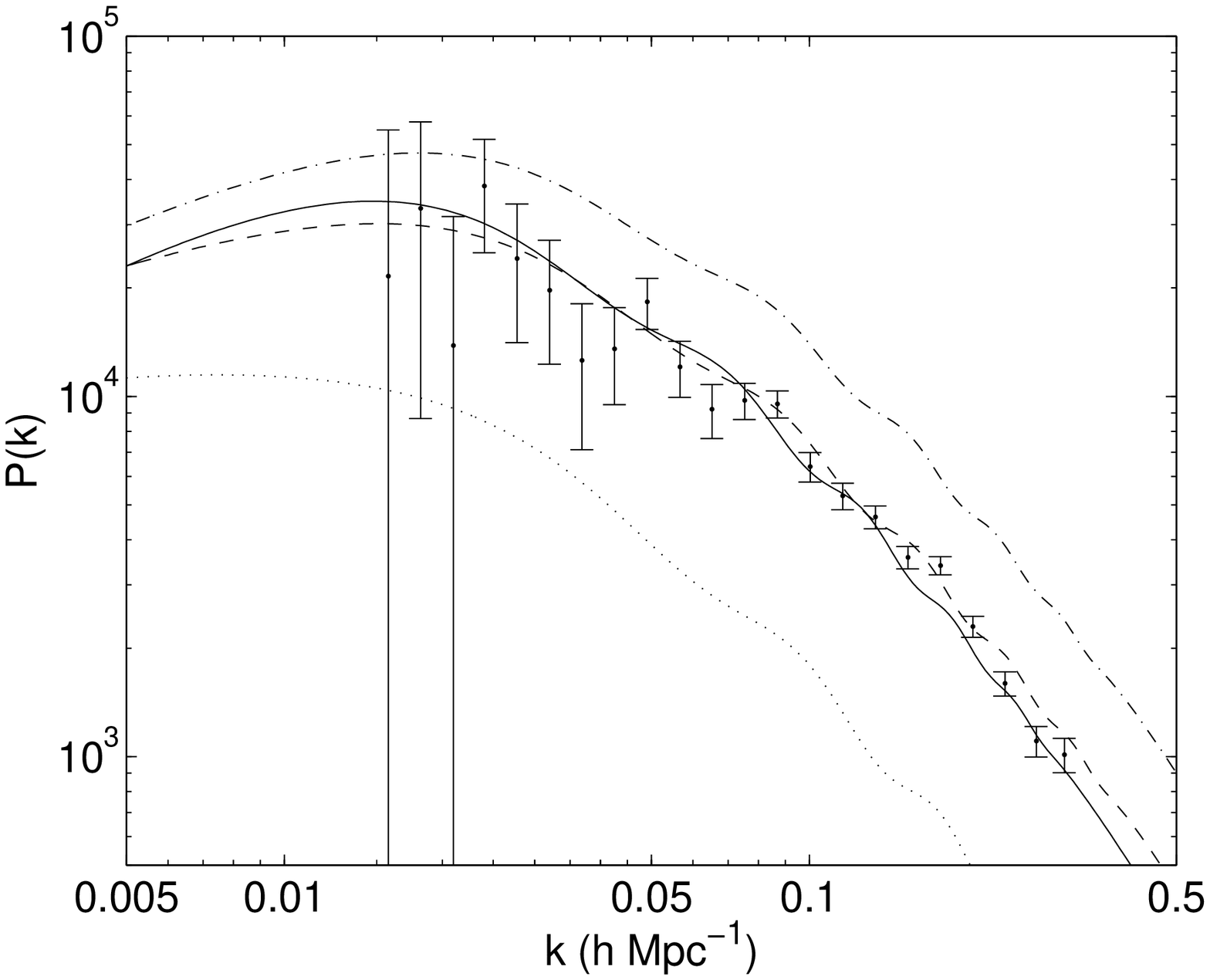}
\includegraphics[width=0.35\textwidth]{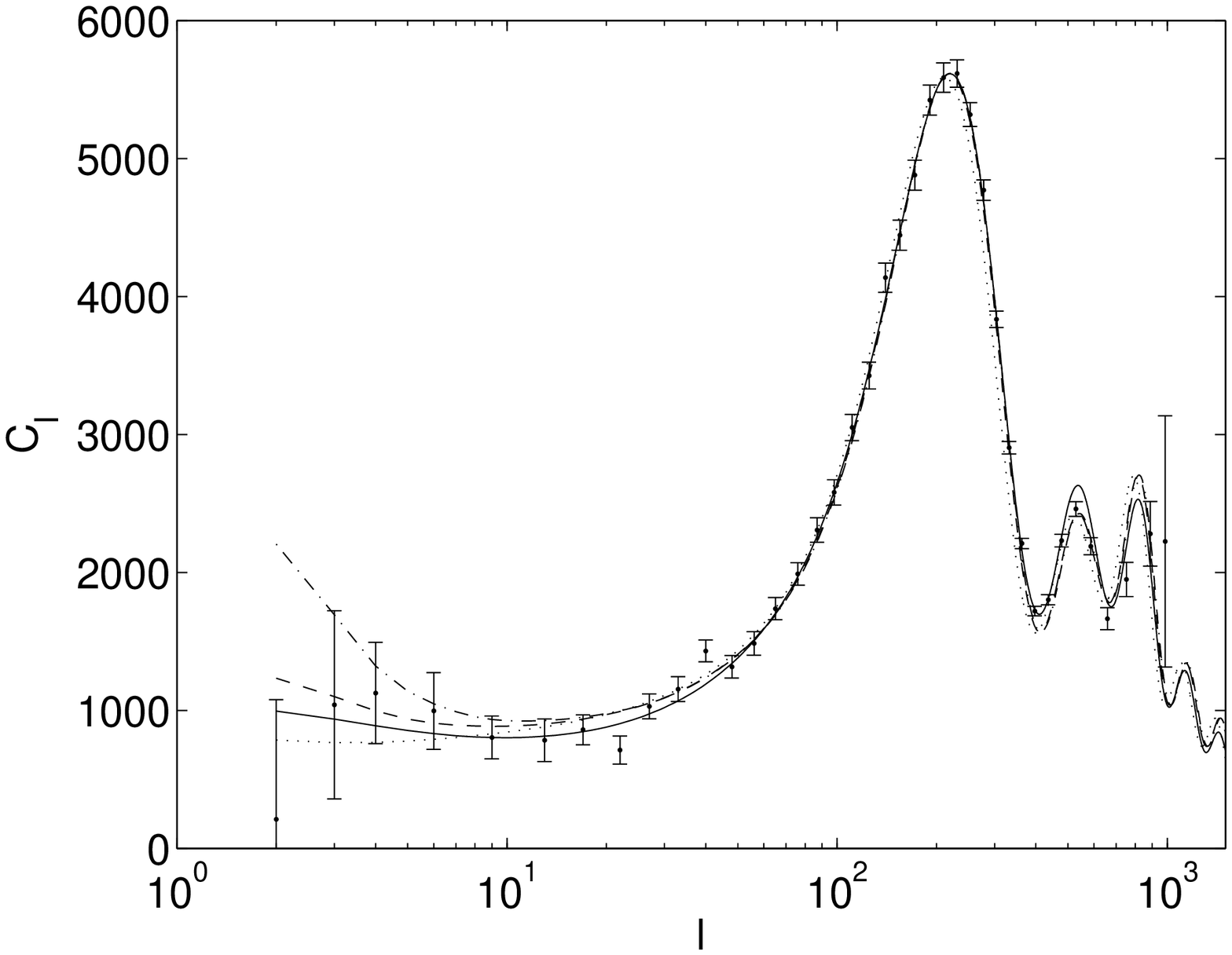}
\includegraphics[width=0.35\textwidth]{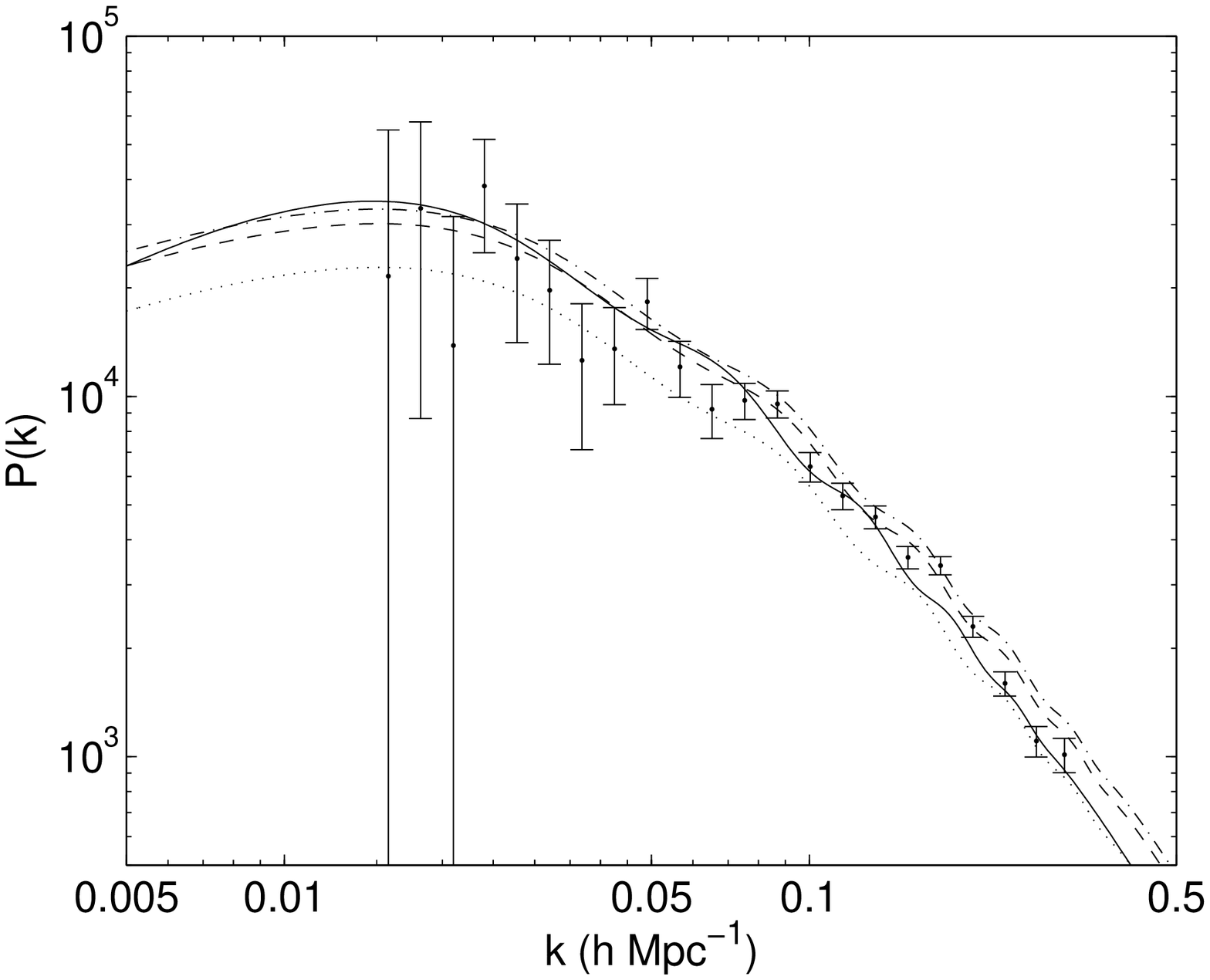}
\caption{\label{figs_2} Top two figures: The effect of the potential slope on the CMB and matter power spectra.
Here $\alpha=20$. Dotted lines are for $\lambda=4.5$, dashed line for $\lambda=6.0$,
and dash-dotted for $\lambda=8.0$.  Bottom two figures: The effect of the coupling slope on the CMB and matter power spectra.
Here $\lambda=6.0$. Dotted lines are for $\alpha=10$, dashed line for $\alpha=20$,
and dash-dotted for $\alpha=30$. The solid line is $\Lambda$CDM model. $\Omega_m^0=0.4$ for all figures. The CMB and matter power spectra error bars are from the  
WMAP data \cite{Spergel:2006hy} and  SDSS\cite{sloan} respectively.}
\end{center}
\end{figure}
In FIG. \ref{figs_2} we show how the cosmological predictions
are changed when the slope of the potential or of the coupling are varied. The main imprint from different
potential slopes $\lambda$ seems to be in the normalization. For low values of $\lambda$, there is significant
contribution of the scalar field during the matter dominated era. This slows
down the rate of growth of matter inhomogeneities. Hence the fact that 
there is less structure nowadays than for larger $\lambda$ is not a 
consequence of the GB 
modification, but rather an effect of the presence of the scalar field
in the earlier scaling era. Finally, in the bottom of FIG. \ref{figs_2} we see that the strength of the 
coupling $\alpha$ might be more difficult to deduce from these data. With steep coupling slopes, the
scalar field domination takes place more rapidly and with more negative $w_{eff}$, which can somewhat amplify
the ISW effect. The contrary happens for smaller $\alpha$.

\begin{figure}
\begin{center}
\includegraphics[width=0.45\textwidth]{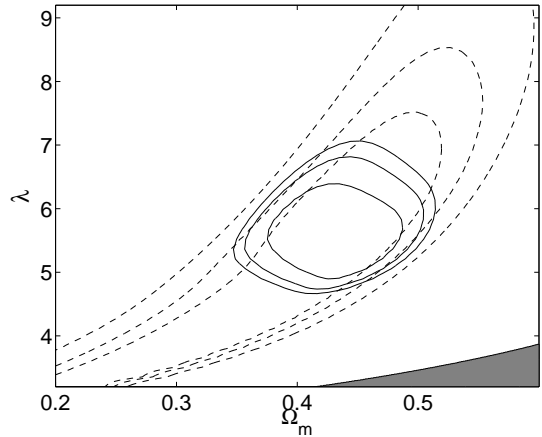}
\caption{\label{contours1} The 68, 90 and 99 percent confidence limits
for the model in the $\Omega_m$ - $\lambda$ plane when $\alpha$ is marginalized
over in the range $1.5\lambda < \alpha <10\lambda$. Dotted lines are constraints from
the SNeIa data and the solid lines from the combined SNeIa and CMBR shift 
parameter data. In the gray area the
scaling solution is unstable regardless of $\alpha$.
If the scalar field is tracking already in the nucleosynthesis
epoch, there is a tension with the amount of early quintessence and the nucleosynthesis
limit. Conservatively, this limit would translate to  $\lambda>6.3$ \cite{Copeland:2006wr}. }
\end{center}
\end{figure}

We now consider the
constraints arising from astrophysical and cosmological observations. We 
calculate the SNeIa luminosity-distance relation. To compare with data, we 
use the ''Gold'' sample of 157 SNeIa from 
Ref.  \cite{Riess:2004nr} and marginalize over the Hubble constant $H_0$. We also compute the CMBR 
shift parameter \cite{Odman:2002wj} $\mathcal{R}$ and apply the latest constraints \cite{Wang:2006ts}. 
The combined constraints arising from all these data is shown in FIG.\ref{contours1}. The SNeIa data
alone is consistent with a wide range of matter densities $\Omega_m$, but when combined with 
the CMBR parameter $\mathcal{R}$ one is restricted to rather high $\Omega_m \sim 0.4$. Note that we restrict here 
only to cosmologies where a scaling matter era is followed by the acceleration era. The existence of the scaling 
attractor requires $\lambda>\sqrt{6}$, otherwise the evolution is different and depends on the initial conditions 
for the field. Therefore we leave models with lower potential slopes, $\alpha<\sqrt{6}$, out from this study, but 
their phenomenology could be interesting to investigate elsewhere\footnote{Such potentials are in accordance with 
various compactification schemes and particle physics models. One also notes that then $\alpha>\lambda$ is not 
required for the transition to acceleration.}.  

Time variation of the effective gravitational constant $G_*$ 
is tightly constrained by observations within the Solar system and laboratories, and indicate that
$|G_*'/G_*| \lesssim 0.01$  \cite{Uzan:2002vq}. 
To derive the variation of this constant for the coupled GB
gravity, we follow the approach of Ref. \cite{Amendola:2005cr} where
cosmological perturbation equations were considered at their Newtonian
limit. 
The
Poisson equation was derived and the effective strength of gravitational
coupling was read from the resulting expression, which relates the gradient of the
gravitational potential to the perturbations in matter density. The result in our
case is
\be \label{g_eff}
\frac{G_*}{G} = 4\frac{-x^4 + \mu^2(1+\epsilon)^2 +
        x^2[2(1 + \epsilon)(\mu-1) + y]}{x^2[4 + \mu(5\mu -8)] - \mu^2[6(1 + \epsilon)(\mu-1) + y]},
\ee
consistently with Ref. \cite{Amendola:2005cr}. 
Though not obvious from the formula, it equals one when the coupling
goes to zero. Generally, when $\mu$ is of order one, then one expects the $G_*'/G_*$ to be 
of roughly of order one as well. As claimed in Ref.  \cite{Amendola:2005cr}, one has to 
assume ''an accidental cancellation'' to satisfy the bound for $G_*'/G_*$ in the presence of
significant GB contribution to the energy density. 
\footnote{
Such a tight bound might not be so problematical if one takes into account that 
cosmological variations of $G$ and other gauge-couplings might be different from the ones we
measure on Earth or within our Solar system  \cite{clifton}.}.
This means that we have to fine-tune the coupling parameter $\alpha$ in order to eliminate
the time variation of the effective gravitational coupling, typically with the accuracy of $0.01$. 
Then the Newtonian limit in general exhibits time-varying $G_*$, but at present days
this $G_*$ appears to us as a constant. We have reported the cosmological
results in such a case in Table \ref{tabu}.  The best-fit values of $\chi^2$ per 
effective degree of freedom, $\chi^2_{dof}$, are slightly better than in the $\Lambda$CDM
case.
\begin{table}
\center
\label{tabu}
\begin{tabular}{|c||c|c||c|c|c|c|}
\hline
Data set  & \multicolumn{2}{|c||}{$\Lambda$CDM} & \multicolumn{4}{|c|}{$R^2_{GB}\phi$ model}  \\
\hline
          & $\chi^2_{dof}$ & $\Omega_m$ & $\chi^2_{dof}$ & $\Omega_m$ & $\lambda$ & $\alpha$ \\
\hline
 SNeIa     & $ 1.142 $  & $ 0.314 $    &  1.146   & 0.42      & 5.1
&   32.3     \\
\hline
 SNeIa+$\mathcal{R}$   & $ 1.144 $  & $ 0.277 $    &  1.141   & 0.44       
&  5.2
&   33.8     \\
\hline
\end{tabular}
\caption{Best-fit values for
$\Lambda$CDM model compared with fits of the coupled scalar field model
for some parameter values when the coupling $\alpha$ is set in order
to fix the present time variation of $G_*$ to zero. The degrees of freedom
in the first row are $157-d$, and in the second $158-d$, where
$d=1$ for the $\Lambda$CDM and $d=3$ for the $R^2_{GB}\phi$ models.}
\end{table}

Interestingly, $G_*$ determines the cosmological evolution of
matter inhomogeneities at subhorizon scales. 
The matter overdensity $\delta$ evolves according to \cite{we}
\be \label{d_evol}
\ddot{\delta} + 2H\dot{\delta} = 4\pi G_*\rho \delta.
\ee
From this equation we can make the important conclusion that the evolution of
matter inhomogeneities at small scales is independent of wavelenght. Thus we expect, 
given the same primordial spectrum of perturbations, the main difference in the matter 
power spectrum (when compared to $\Lambda$CDM) to be in the normalization. Note that
to arrive at this result, we have not assumed specific forms for the coupling or the
potential, but only that the scalar appearing in the action (\ref{action}) is not 
very massive. Our numerical solutions of the whole perturbation system confirm that 
the approximation Eq.(\ref{d_evol}) indeed is good at subhorizon scales. 

The effective gravitational constant $G_*$ can in principle diverge. We
find that in the example model presented here, $G_*$ typically diverges in the future, and for low matter
densities this can happen even before $a=1$. It is unclear what happens at such
point, since the linear approximation certainly breaks down near the (what would be) singularity.
Matter perturbations will at least for a while grow explosively. It is possible that
consequently the de Sitter phase will not be then reached,
which could help to define the S-matrix in string theory.
This divergence can be related to previous stability considerations 
of these models (for recent studies, see Refs.\cite{DeFelice:2006pg,Calcagni:2006ye}). 
This far stability conditions for these models have been derived only in the case that
the field perturbations decouple from other fluids.
However, in dark energy cosmologies
such as here one should take into account both matter and the corrections to Einstein gravity. 
We provide a clear indication that the scalar action in its vacuum 
form determines the stability also in the case that $\rho_m \neq 0$. 
The 
action for the potential $\Phi$ in vacuum features an effective propagation speed \cite{Hwang:2005hb}
\be
s_{SC} = \frac{-x^2[4+\mu(5\mu-8)]+\mu^2[6(1+\epsilon)(\mu-1)+y]}{(\mu-1)[3\mu^2-4(\mu-1)x^2]},
\ee
from which we can see that the linearized matter perturbations diverge exactly at the
points where this propagation changes its sign \footnote{An occurrence 
of $w_{eff}<-1$ does not have a direct relation with these conditions. Hence it is 
possible to have phantom universe and avoid the divergence of perturbations.}.

To summarize, we studied a string-inspired low energy action where a coupling between a scalar field and the 
GB invariant is present. 
Solving the full system of equations which describe both the homogeneous universe as well as its first order 
perturbations
we found several new insights to GB dark energy cosmologies in general, which could be relevant to future 
investigations of viable 
string-theory low energy effective actions. 
We showed that the background universe can present an attractor solution which features 
late transition from a scaling era to acceleration triggered by the GB coupling. This mechanism
might be seen to alleviate the coincidence problem. In addition, the model can consistently explain a presently
ongoing but transient phantom era. This background expansion
can exhibit a good a accordance with cosmological and astrophysical data. 
%
The evolution of matter perturbations is scale-invariant at small
scales in the presence of the GB term, and thus the shape of the matter
power spectrum is retained. 
Hence, the latest data from the CMBR anisotropies as well as LSS can 
show agreement with these models. 
However, using a combined set of present days cosmological observations it is possible to 
constrain the parameters of the theory tightly. 


\section*{Acknowledgments}
DFM acknowledges support from the Research Council of Norway, 
through project number 159637/V30,
and from the
Perimeter Institute where part of this work was undertaken. 
TK is supported by the Magnus Ehrnrooth Foundation.
%
%

\end{document}